\documentclass[a4paper,notitlepage,12pt]{article}

\begin{document}
\title{Multiscale simulation on shearing transitions of thin-film lubrication with multi-layer molecules}
\author{Z.-B. Wu\footnotemark[1]$^1$ and X. C. Zeng$^2$\\
$^1$LNM, Institute of Mechanics, Chinese Academy of Sciences,\\
 Beijing 100190, China; School of Engineering Science, \\
 University of Chinese Academy of Sciences, Beijing 100049, China\\
$^2$Department of Chemistry, University of Nebraska-Lincoln,\\
Lincoln, NE 68588, USA}
\maketitle

\footnotetext[1]{Corresponding author. Tel:. +86-10-82543955;
fax.: +86-10-82543977. \\
Email addresses: wuzb@lnm.imech.ac.cn (Z.-B. Wu)}
\newpage

\begin{abstract}
Shearing transitions of multi-layer molecularly thin-film lubrication systems
in variations of the film-substrate coupling strength and the load
are studied by using a multiscale method. Three kinds of the interlayer slips
found in decreasing the coupling strength are 
in qualitative agreement with experimental results.
Although tribological behaviors
are almost insensitive to the smaller coupling strength,
they and the effective film thickness
are enlarged more and more as the larger one increases.
When the load increases, the tribological behaviors are
similar to those in increasing coupling strength, but the effective
film thickness is opposite.
\\



\end{abstract}
\newpage

\section{Introduction}

A thin fluid film confined between two solid substrates
may prevent surfaces of the substrates coming into molecular contact and becoming damaged in the dry friction,
which is termed as thin-film lubrication.
The systematic friction force is reduced three orders of magnitudes when the dry friction is replaced by the
thin-film lubrication. The thin-film lubrication is a
very interesting topic on both fundamental theory and engineering application, such as 
viscous dampers, magnetic storage devices and designs of MEMS/NEMS\cite{1,2,3}.
A stick-slip phenomenon (the freezing-melting transition) appears in
the molecularly thin film lubrication when the substrates are assumed as rigid.
The critical friction force, which is quantized with the number of fluid molecular layers separating the surfaces,
decreases as the number of fluid molecular layers increases\cite{3d,3e,ss,ds}.
For the thin film lubrication with multi-layer fluid molecules,
the slipping process may have the partial and whole film melting due to the different interaction potentials between
the solid and fluid molecules\cite{a,b}.
Although the surfaces of the substrates are smooth on macroscopic scale, they
actually make molecular contact through the medium of the thin-film.
In particular, the solid substrates may have multiple spatial scales and influence
the microscopic motions of both the solid and fluid molecules near/within the film\cite{3a,3b,3c}.
It is therefore necessary to include the details of molecular motions in the substrates
to treat the shearing transitions of the thin-film lubrication with multi-layer fluid molecules.
To fully understand the physical mechanism of the thin-film lubrication system,
the heterostructure interface is described by an atomistic model,
while the elastic behavior of the much thicker substrate is modeled by the finite element method.
Such multiscale problem of the thin-film lubrication system must be reflected in a theoretical model.

During the past three decades a series of effective methods combining atomistic and continuum descriptions
of fluids and solids are developed\cite{4,5,6,6a,7,7a,8,9,10,11,11a,12,13,14,15,16,17,18,19,20,21,22}.
On the one hand, several effects on fluid flow near the walls have been made by applying the molecular
dynamics to follow the trajectories of the fluid atoms in the layers near the walls and extend the non-sliding
boundary conditions near the walls to the sliding ones. Beyond that the hydrodynamic equations are employed,
the molecular dynamics and hydrodynamic treatments are self-consistently joined in the transition region
that overlaps the two regimes of the atomistic and continuum descriptions.
On the other hand, within the context of solid materials, two kinds of the effects to combine atomistic
and continuum descriptions of the behaviors of materials have been completed at zero temperature.
One is the FEAt procedure, where
a core region described at the atomistic scale is surrounded by a transition region connected with
the continuum regions covered by finite elements(FE). The other is a quasicontinuum treatment, where
the whole lattices of material are overlayed with a FE mesh. The energy density in each FE is evaluated
with the movements of its nodes. By including effects of thermal motions on the dynamics of certain processes,
the FEAt and quasicontinuum methods are extended to finite temperature. Moreover, in a treatment
of crack propagation in pure silicon, to make a multiscale modeling approach that dynamically couples different length scales
and accounts for thermal effects,
 the macroscopic atomistic ab-initial dynamics(MAAD) is developed. The molecular dynamics(MD) with the tight-binding approximation,
 the MD with the semi-empirical potential energy and the FE method are applied in the different length scales
 ranging from the atomic scale through the microscale and finally to the mesoscale/macroscale.
 In particular, the soft matter as a subfield of condensed matter comprising a variety of biological systems
 such as macromolecular assemblies has both fluid and solid properties.  To investigate its long time- and length-scale behaviors,
  the multiscale approaches with quasi-equilibrium assumptions\cite{16a,16b}, the hybrid continuum method mixing the classical
 force field and coarse grained model\cite{16c,19a} and the multiscale algorithm with the order parameters
 \cite{18a,20a} have been developed. The multiscale analysis as a powerful methodology plays a critical role
 in the biological processes.

It is noted that for the thin-film lubrication system, the evolution of the system is slower
than the molecular motions. In other words, on the time scale of the relevant process the system
remains thermodynamic equilibrium. Such reversible or quasi-static processes of the thin-film lubrication
can be described by means of Monte Carlo methods.
In previous articles\cite{24,25} we developed the multiscale description of fluid-solid interfacial systems
and applied it to treat the monolayer film lubrication with elastic substrates. The far-region solid substrates are coarse-grained
by local and nonlocal elements. The systematic free-energy related to the elements is corrected by the local harmonic approximation.
The principal conclusion of
papers is that the hybrid atomistic-coarse-graining (HACG) scheme yields the shear-stress profiles
and the mean separation curves in good agreement with those in the fully atomistic description of the
system over a wide variety of conditions. However, since the simulations
were performed only for a monolayer molecularly film lubrication, an extension to multi-layer molecularly
film lubrication systems and an investigation of microscopic structures during the shearing slips
are the purpose of the present article.
In particular, the tribological behaviors in the multi-layer molecularly film lubrication system not only depend on the elasticity of
substrates, but also on the adhesion strength between the film and substrates and the cohesive strength of the film.
It was shown that an optimal choice of the interaction within fluids relative to the interaction between the fluids and substrates
may reduce the kinetic friction force of the system\cite{c}.
Therefore, it is important to investigate the effects of the adhesion strength between the film and substrates on the
shearing transitions in the system when the cohesive strength of the film is fixed.

The paper is organized as follows. In section 2, a short description
of the coarse-graining model and the computational procedure is provided.
 The numerical results for two cases of three-layer and four-layer molecularly film lubrication
 are analyzed in section 3. Finally, in section 4, some conclusions and discussions are given.

\section{Coarse-graining model and computational procedure}

The idealized 2D contact consists of two identical hexagonal close-packed
crystalline substrates separated by a multi-layer molecularly thin-film at an atomically flat
interface, as shown schematically in Fig. 1. The top/bottom wall is taken
to be rigid with the nearest-neighbor distance, i.e.,  the
lattice constant ${\it a}$.
The bottom wall remains stationary in the
``laboratory'' reference frame; the top wall can be translated
in the $x$- and $y$- directions, but remains
parallel with the bottom wall. The walls serve as handles
by which the substrates can be manipulated.
The lateral alignment of the walls is specified by the register $\alpha$, which is defined by
\begin{equation}
x^t_i=x^b_i + \alpha a,
\end{equation}
where $x^t_i$ u and $x^b_i$ denote the corresponding
atomistic positions in the top and bottom walls, respectively.
$\alpha$ is the fraction of the lattice constant by which the top wall is displaced
laterally with respect to the bottom wall.

The tribological system only comprises the solid substrates
plus the fluid film, but does not cover the walls.
The solid substrates are divided into the near- and far-regions,
which are depicted by using the atomistic and coarse-grained
descriptions, respectively.
The coarse-grained far regions of substrates are covered with
 a mesh of triangular elements.
This coarse-graining partitions the
original substrate atoms into two subsets: $N_n$ nodal atoms
and $N_s$ non-nodal atoms. If one integrates the Boltzmann
factor over the 2$N_s$ degrees of freedom of the non-nodal
atoms, one obtains an effective potential energy governing
the motion of just the nodal atoms
\begin{equation}
V_{eff}({\bf R}^{N_n}) = \sum_{e=1}^{N_e} (N_a^e \tilde{u}_e +N_s^e f_e),
\label{2.8}
\end{equation}
where ${\bf R}^{N_n}$ stands for the nodal configuration, $N_e$ for
the number of elements, $N_a^e$ the number of atoms
underlying element $e$. The configurational energy per atom $\tilde{u}_e$
is expressed as
\begin{equation}
\tilde{u}_e = \frac{1}{2} \sum_{j \neq i} u_{ss}(r_{ij}),
\label{2.9}
\end{equation}
where $i$ denotes the ``centroid'' atom (i.e., the atom nearest
the centroid of $e$) and $j$ labels atoms that lie within the circle
of radius $r_c$ that is centered on $i$.
 The following shifted Lennard-Jones (12,6) potentials are taken as the pair interactions
of the atoms
\begin{equation}
u_{ab}= \left \{ \begin{array}{ll}
		\phi_{ab}(r) - \phi_{ab}(r_c), & \mbox{if}~~~ r<r_c;\\
		0,	& \mbox{if}~~~ r \ge r_c,
		\end{array}
		\right.
\end{equation}
where
\begin{equation}
 \phi_{ab}(r)= 4 \epsilon_{ab} [(\sigma/r)^{12} - (\sigma/r)^6], \,\,\,
ab=ff, fs, ss.
\end{equation}
The effective diameter $\sigma$, $\epsilon_{ab}$ and range $r_c$ is the same for all
pairs. $f_e$ is the Helmholtz energy per atom.
By using the local harmonic approximation\cite{lesar}, $f_e$ can be estimated as
\begin{equation}
f_e = 2k_BT {\rm ln} [h ({\rm det} D)^{1/4} /k_BT],
\label{2.9a}
\end{equation}
where the elements of the $2 \times 2$ dynamical matrix are given by
$D_{kl} =m^{-1} (\partial^2 \tilde{u}_e/\partial x_k \partial x_l)_0 (k, l=1,2)$,
$k$ and $l$ label Cartesian components $(x_1=x, x_2=y)$ of the position
of the reference atom, and the subscript 0 signifies that the partial
derivative is evaluated at the equilibrium configuration.
To show the dynamic properties of molecularly thin-film lubrication, it was found that
 the numerical results based on the Lennard-Jones potentials
are in qualitative agreement with the experimental ones\cite{3d}.

The total configurational energy $U_{cf}$ and
the shear stress $T_{yx,cf}$
of the free-energy corrected HACG system are respectively given
\begin{equation}
\begin{array}{ll}
U_{cf}({\bf r}^{N_f}, {\bf r}^{N'_s},{\bf R}^{N_n}) = &
\frac{1}{2} \sum_{i=1}^{N_f} \sum_{j \neq i}^{N_f} u_{ff}(r_{ij})
 + \sum_{i=1}^{N_f} \sum_{j=1}^{N'_s} u_{fs}(r_{ij})\\
&+ \frac{1}{2} \sum_{i=1}^{N'_s} \sum_{j \ne i}^{N'_s} u_{ss}(r_{ij})
+ \frac{1}{2} \sum_{i=1}^{N'_s} \sum_{j=1}^{N''_s} u_{ss}(r_{ij})\\
&+ \frac{1}{2} \sum_{i=1}^{N_w} \sum_{j=1}^{N''_s} u_{ss}(r_{ij})
+ \sum_{e=1}^{N_e} N_a^e [\frac{1}{2} \sum_{j \neq i} u_{ss}(r_{ij})]\\
&+ \sum_{e=1}^{N_e} N_s^e 2k_BT {\rm ln} [\hbar ({\rm det}D)^{1/4} /k_BT]
\label{2.10}
\end{array}
\end{equation}
and
\begin{equation}
\begin{array}{ll}
T_{yx,cf} =& \frac{1}{2L_x} \sum_{i=1}^{N_f} \sum_{j \neq i}^{N_f}
< u'_{ff}(r_{ij}) x_{ij} y_{ij} / (r_{ij} L_y) >\\
&+ \frac{1}{L_x} \sum_{i=1}^{N_f} \sum_{j=1}^{N'_s}
< u'_{fs}(r_{ij}) x_{ij} y_{ij} / (r_{ij} L_y) >\\
&+ \frac{1}{2L_x} \sum_{i=1}^{N'_s} \sum_{j \neq i}^{N'_s}
< u'_{ss}(r_{ij}) x_{ij} y_{ij} / (r_{ij} L_y) >\\
&+\frac{1}{2L_x} \sum_{i=1}^{N'_s} \sum_{j=1}^{N''_s}
< u'_{ss}(r_{ij}) x_{ij} y_{ij} / (r_{ij} L_y) >\\
&+\frac{1}{2L_x} \sum_{i=1}^{N_w} \sum_{j=1}^{N''_s}
< u'_{ss}(r_{ij}) x_{ij} y_{ij} / (r_{ij} L_y) >\\
&+\frac{1}{L_x} \sum_{e=1}^{N_e} N_a^e [\frac{1}{2} \sum_{j \neq i}
< u'_{ss}(r_{ij}) x_{ij} y_{ij} / (r_{ij} L_y) >]\\
&+\frac{1}{L_x} \sum_{e=1}^{N_e} N_s^e [\frac{k_BT}{2}
<\frac{1}{|D|} \frac{\partial |D|}{\partial {\alpha a}}>],
\label{2.12}
\end{array}
\end{equation}
where $N'_s$ stands for the number of atoms in the near regions of the
substrates, $N''_s$ for the number of
underlying atoms in the far regions,
$N_e$ for the number of elements, $N_a^e$ ($N_s^e$) the number of atoms (non-nodal atoms)
underlying element $e$,
$k_B$ is Boltzmann¡¯s constant, $\hbar$ is Planck¡¯s constant
and $D$ is the local dynamical matrix associated with the centroid atom in a local element.
The sliding with the register $\alpha$ is viewed as a quasistatic (or reversible) process.
To compute the thermomechanical properties in the forward and backward processes we perform isothermal-isobaric
Monte Carlo (MC) simulations.
More details of the numerical method were
presented in \cite{24,25}.

For the simulations, Table I lists the values of the various
parameters in the system. Numerical values are expressed in dimensionless
units based on the Lennard-Jones parameters for the
solid-solid interaction: distance is expressed in units
of $\sigma$; energy in units of $\epsilon_{ss}$; stress
in units of $\epsilon_{ss}/\sigma$; temperature in
units of $\epsilon_{ss}/k_B$. The film-substrate attractive well depth is supposed to
obey a modified combining rule $\epsilon_{fs} = \kappa \sqrt{\epsilon_{ff}\epsilon_{ss}} =\kappa/3$, where
$\kappa$ is an adjustable constant. $\epsilon_{fs}$ and $\epsilon_{ff}$ represent approximately
the adhesion strength between the film and substrates and the cohesive strength of the film in the system, respectively.
In the following, we take two cases of (three-layer and four-layer) molecularly thin-film to investigate
effects of the film-substrate coupling strength and loads on
the tribological dynamical processes of the thin-film lubrication system
at fixed $\epsilon_{ss}$ and $\epsilon_{ff}$.

\section{Results}
\subsection{Three-layer molecularly thin-film lubrication}
\subsubsection{Effects of film-substrate coupling strength at a fixed load}

Figure 2 displays plots of the shear stress $T_{yx}$ and the mean separation $L_y$  between walls
for the three-layer molecularly thin-film lubrication system at the load $T_{yy}$=-1.0 as a function
of register $\alpha$ for different $\kappa(\epsilon_{fs})$.
Since all shear-stress and mean separation profiles possess similar behaviors, we take $\kappa$=1 ($\epsilon_{fs}$=1/3)
as an example to summarize their common
features.  The profiles display a forward periodic process in $\alpha$
with a period length 1, which composites
two elastic (stick) regions ($0 \leq \alpha <0.6$, $0.7 < \alpha \leq 1$)
and a transition (slip) region ($0.6 \leq \alpha \leq 0.7$).
At $\alpha =0$, fluid atoms with an initial random distribution converge to a stable close-packed
structure with solid atoms of the upper and lower substrates as shown in Fig. 3(a).
The perfect symmetric atomistic structure exhibits the system
in an equilibrium state and leads to $T_{yx,cf}$=0.
As $\alpha$ increases gradually in the range $0 < \alpha < 0.6$,
the upper and lower substrates adsorbing the film just like a whole crystal
to generate an elastic shear behavior in the $x$ direction.
The film atoms as in a solid state hold the shear stress in the system.
A net force provided to the
upper substrate/the film by the film/the lower substrate varies stronger
and leads to the increase of $T_{yx,cf}$. Meanwhile, to hold the shear stress
in the thin film, the effective film thickness (or the mean separation  $L_y$) also increases.
As $\alpha$
varies in the range $0.6 \leq \alpha \leq 0.7$, the systematic shear strain reaches a critical value,
which corresponds to the largest shear stress.
For the lubrication system, the interaction between the film
and the substrates or among the interlayers in the film cannot hold the
strong shear stress. Under the action of the shear stress, the interlayer atoms in the film
or the surface atoms of the upper substrate
abruptly stride over the atoms in the layer touching with them as barriers,
so that the upper and lower substrates make a relative
slippage to restrain the increase of the systematic shear stress.
After the sliding process, the system performs again an elastic shear behavior.
Since the net force
provided to the upper substrate/the film by the thin film atoms/the lower substrate has changed
its direction, the shear stress is negative.
In this process, the effective film thickness (or the mean separation $L_y$)
reaches the largest value and then drops gradually.
When $\alpha$
increases continuously in the range $0.7 < \alpha <1$,
the systematic behavior is asymmetric with that at $1-\alpha$. When the register
finally reaches $\alpha=1$, the system is precisely in
the same state as it is at $\alpha=0$. As $\alpha$ increases
from 1 to 1.2, the above profiles in the range $0 < \alpha \le 0.2$ are repeated.
Shearing from $\alpha=1.2$ to $\alpha=0$ carries the system
in reverse through the same states as shearing in
the forward direction. In the forward and backward processes,
the slip process deviates from the center $\alpha=0.5$ but appears at its both sides
duo to the longer relaxation time required to reach the quasi-equilibrium state\cite{26,27}.
So the paths for the stick and the slip processes constitute a hysteresis loop.

In Fig. 2(a), the thermal elastic coefficient ($c_{\kappa}=\frac{\Delta T_{yx,cf}}{\Delta \alpha}$) in the stick process
decreases when $\kappa$ decreases from 3 to 1/6. It is almost a constant when $\kappa$ decreases
over $1/6 \le \kappa \le 1/20$. The thermal elastic coefficient reflects the ability of the system resisting the shear strain.
The larger $\kappa (\epsilon_{fs})$ has influence on  $c_\kappa$ while
the smaller $\kappa (\epsilon_{fs})$ is insensitive to it.
So only the larger one between  $\epsilon_{fs}$ and $\epsilon_{ff}$ is more significant
to determine the thermal elastic coefficient of the system.
Meanwhile, the width of the hysteresis loop reveals the deviation of the system from the quasi-equilibrium state.
The larger $\kappa (\epsilon_{fs})$ corresponds to the more width of the hysteresis loop while
the smaller $\kappa (\epsilon_{fs})$ is insensitive to it.
So the width of hysteresis loop has the above similar features with the thermal elastic coefficient.
Moreover,
in Fig. 2(b), the mean separation $L_y$ (or the effective film thickness) monotonically decreases with decreasing $\kappa$
except $\kappa$=3.
In the interaction between the film and the substrates
at $\kappa$=3 ($\epsilon_{fs}=\epsilon_{ss}$),
the film atoms adsorbed by the substrates
take effects as the surface atoms of the substrates. So the effective film thickness is reduced approximately from three-layer
to monolayer.  This leads to the decrease of $L_y$.
In order to display the effects of $\kappa$ on the shearing transitions, three kinds of
the atomistic configuration of the film and the near substrates at $\alpha=0$
and $\alpha=1.0$ for the different $\kappa(\epsilon_{fs})$ are shown in Fig. 3.

(1) The slip process appearing in the internal layers of the film atoms is termed as the internal slip.
(i)For $\kappa=3$ and 1, $\epsilon_{fs}$(=1 and 1/3) is larger than $\epsilon_{ff}$.
In the thin-film lubrication system, the weakest interaction among atoms
appears in the film.
The slip process happens between the first and second layers in the film as shown in Fig. 3(b).
The fluid atoms of the first layer are moved a lattice distance to the right with respect to
the second layer in the film.
At the same time, the upper substrate adsorbing the fluid atoms of the first layer
is moved together with the first layer in the film.
(ii) For $\kappa=1/3$, $\epsilon_{fs}$(=1/9) is equal to $\epsilon_{ff}$.
In the thin-film lubrication system, the weaker interaction of the atoms appears in the film
and between the substrates and the film.
The slip process happens between the second and the third layers in the film as shown in Fig. 3(c).
The fluid atoms of the second layer is moved a lattice distance to the right with respect to
the third layer in the film.
In this case, the first layer in the film adsorbed by the upper substrate is moved together with the second layer
in the film.

(2) The slip process appearing between the film and substrates is termed as the interfacial slip.
(iii) For $\kappa=1/6$ and 1/9, $\epsilon_{fs}$(=1/18 and 1/27) is smaller than $\epsilon_{ff}$.
In the thin-film lubrication system, the weakest interaction among atoms
appears between the substrates and the film.
 The slip process happens between the upper substrate and the first layer in the film as shown in Fig. 3(d).
The surface atoms of the upper substrate are moved a lattice distance to the right with respect to the first layer
in the film.
(iv) For $\kappa=1/12$, $\epsilon_{fs}$(=1/36) is smaller than $\epsilon_{ff}$.
In the thin-film lubrication system, the weakest interaction among atoms also
appears between the substrates and the film.
 The slip process happens between the third layer in the film and the lower substrate as shown in Fig. 3(e).
The fluid atoms in the third layer are moved a lattice distance to the right with respect to the lower substrate.

(3) The slip processes appearing not only between the internal layers of the film atoms but also between the film and substrates
are termed as the mixing slip, i.e.,  mixing of the internal and the interfacial slips.
(v) For $\kappa=1/16$ and 1/20, $\epsilon_{fs}$(=1/48 and 1/60) is smaller than $\epsilon_{ff}$.
In the thin-film lubrication system, the weakest interaction among atoms
appears between the substrates and the film.
 The slip processes happen simultaneously between the upper substrate and the first layer in the film,
 between the first and second layers in the film and
 between the third layer in the film and the lower substrate as shown in Fig. 3(f).
 Only the second and the third layers in the film move together.
  The surface atoms in the upper substrate are moved a lattice distance to the right with respect to the first layer in the film.
 The fluid atoms in the second and third layers are moved a lattice distance to the left with respect to
the first layer in the film and the lower substrate.

\subsubsection{Effects of loads at a fixed film-substrate coupling strength}

Figure 4 displays the shear-stress and the mean separation profiles of the three-layer molecularly thin-film lubrication
 at $\kappa=1$ ($\epsilon_{sf}=1/3$) for different load $-T_{yy}$ on the system.
As the load $-T_{yy}$ increases, the thermal elastic constant $c_1$ increases
and the width of the hysteresis loop enlarges. Meanwhile, the vertical movement of the atoms
in the system is more restrained. It leads to the decrease of the effective film thickness (or the mean separation $L_y$).
For $T_{yy}=-0.5$, the slip process happens between the second and third layers in the film.
The fluid atoms of the second layer are moved a lattice distance to the right with respect to those of
the third layer.
At the same time, the first layer in the film adsorbed by the upper substrate is moved together with the second layer
in the film.
For $T_{yy}=-1.0$, the slip process happens between the first and second layers in the film.
The fluid atoms of the first layer are moved a lattice distance to the right with respect to those of
the second layer.
For $T_{yy}=-1.5$, the slip process happens between the second and third layers in the film.
The fluid atoms of the second layer are moved a lattice distance to the right with respect to those of
the third layer.
So the internal slip at the fixed $\kappa=1$ ($\epsilon_{sf}=1/3$) is kept as the load increases.
In comparing with the effects of the film-substrate coupling strength on the system, it is found that
the effects of the load $-T_{yy}$ on the system have similar tribological dynamical behaviors
(the thermal elastic constant and the width of the hysteresis loop),
but are opposite on the effective film thickness(or the mean separation $L_y$) in system.

\subsection{Four-layer molecularly thin-film lubrication}
\subsubsection{Effects of film-substrate coupling strength at a fixed load}

Figure 5 displays plots of the shear stress $T_{yx}$ and the mean separation $L_y$
for the four-layer molecularly thin-film lubrication system at the load $T_{yy}$=-1.0 as a function of register $\alpha$
for different $\kappa(\epsilon_{fs})$.
Since all shear-stress
profiles possess similar behaviors, we take $\kappa$=1 ($\epsilon_{fs}$=1/3)
as an example to summarize their common
features.  The profiles display a forward periodic process in $\alpha$
with a period length 1. The periodic process composites
two elastic (stick) regions ($-0.5 \leq \alpha <0.1$, $0.15 < \alpha \leq 0.5$)
and a transition (slip) region ($0.1 \le \alpha \le 0.15$).
At $\alpha =-0.5$, the film atoms with an initial random distribution converge
to a stable close-packed structure with the surface atoms of the upper and lower substrates
as shown in Fig. 6(a).
It exhibits the system in an equilibrium state with $T_{yx,cf}=0$.
As $\alpha$ increases gradually from -0.5 to 0.1, the elastic shear behavior to the $x$ direction
in the combination of the upper and lower substrates connected by the film is enhanced.
This leads directly to the increase of $T_{yx,cf}$. As $\alpha$ increases in the range
$0.1 < \alpha <0.15$, the fluid atoms in the film abruptly stride over
the atoms in the layer touching with them as barriers, so that the
shear strain between the upper and lower substrates are varied from a positive value to a negative
one.  As $\alpha$ increases gradually from 0.15 to 0.5,
 the systematic behavior is asymmetric with that at $-\alpha$,
i.e., the shear strain of the system is varied from the negative value to zero.
 When the register finally
reaches $\alpha=0.5$, the system is in precisely the same state as it is at $\alpha=-0.5$.
As $\alpha$ increases from 0.5 to 0.7, the above scenario ($-0.5 < \alpha <-0.3$) is repeated.
Shearing from $\alpha=0.7$ to $\alpha=-0.5$ carries the system
in reverse through precisely the same states as shearing in
the forward direction.

In Fig. 5, the thermal elastic coefficient $c_\kappa$
decreases when $\kappa$ decreases from 3 to 1/6. It is almost a constant when $\kappa$ decreases
over $1/6 \le \kappa \le 1/20$.
Meanwhile, as $\kappa (\epsilon_{fs})$ decreases, the width of hysteresis loop has the above similar features with
the thermal elastic coefficient.
Moreover, the mean separation $L_y$ (or effective film thickness) monotonically
decreases with decreasing $\kappa$ except $\kappa$=3.
In the interaction between the film and the substrates at $\kappa$=3 ($\epsilon_{fs}=\epsilon_{ss}$),
the film atoms adsorbed by the substrates
take effects as the surface atoms of the substrates, so that the effective film thickness is reduced approximately from four-layer
to two-layer.  This leads to the decrease of $L_y$.
In order to display the effects of $\kappa$ on the shearing transitions, two kinds of
the atomistic configuration of the film and the near substrates at $\alpha=-0.5$
and $\alpha=0.5$ for the different $\epsilon_{fs}$ are shown in Fig. 6.

(1) The internal slip:
(i) for $\kappa=$ 3, 1 and 1/3, $\epsilon_{fs}$(=1, 1/3 and 1/9) is not smaller than $\epsilon_{ff}$.
The slip process happens between the second and third layers in the film as shown in Fig. 6(b).
The fluid atoms of the second layer are moved a lattice distance to the right with respect
to those of the third layer.
In this case, the first layer in the film adsorbed by the upper substrate is moved together with the second layer
in the film.
(ii) For $\kappa=1/6$, $\epsilon_{fs}$(=1/18) is smaller than $\epsilon_{ff}$.
The slip process happens between the third and fourth layers in the film as shown in Fig. 6(c).
The fluid atoms of the third layer is moved a lattice distance to the right with respect to
those of the fourth layer.
At the same time, the second layer in the film is moved together with the third layer
in the film.

(2) The interfacial slip:
(iii) for $\kappa=1/9$ and 1/12, $\epsilon_{fs}$(=1/27 and 1/36) is smaller than $\epsilon_{ff}$.
 The slip process happens between the upper substrate and the first layer in the film as shown in Fig. 6(d).
The surface atoms of the upper substrate are moved a lattice distance to the right with respect to the first layer
in the film.
(iv) For $\kappa=1/12$ and 1/16, $\epsilon_{fs}$(=1/36 and 1/48) is smaller than $\epsilon_{ff}$,
 the slip process happens between fourth layer in the film and the lower substrate as shown in Fig. 6(e).
  The fluid atoms in the fourth layer are moved a lattice distance to the right with respect to
to the lower substrate.

(3) The slip processes appearing not only between the film and the upper substrate but also between the film and the lower substrate
are also termed as the mixing slip, i.e.,  mixing of the upper and lower the interfacial slips.
(v) For $\kappa=1/20$, $\epsilon_{fs}$(=1/60) is smaller than $\epsilon_{ff}$.
The slip processes appear not only between the upper substrate and the first layer in the film
but also between the fourth layer in the film and the lower substrate as shown in Fig. 6(f).
The fluid atoms of the first, second, third and fourth layers are moved a lattice distance to the right with respect to
the lower substrate.
The surface atoms of the upper substrate are moved two lattice distance to the right with respect to the first layer
in the film. So the upper and lower substrates have only the relative slippage with a lattice distance.

\subsubsection{Effects of loads at a fixed film-substrate coupling strength}

Figure 7 displays the shear-stress and the mean separation profiles of the four-layer molecularly thin-film lubrication
 at $\kappa=1$ ($\epsilon_{sf}=1/3$) for different load $-T_{yy}$ on the system.
As the load $-T_{yy}$ increases, the thermal elastic constant increases and
the width of the hysteresis loop enlarges. Meanwhile,
 $L_y$ (or the effective film thickness) becomes thinner.
For $T_{yy}=-0.5$, the slip processes happen not only between the first and second layers,
 between the second and third atomistic layers in the film, but also between the third and fourth layers in the film.
The fluid atoms of the third layer are moved a lattice distance to the right with respect to those of
the fourth layer. The fluid atoms of the second layer are moved a lattice distance to the right with respect to those of
the third layer. The fluid atoms of the first layer are moved a lattice distance to the left with respect to those of
the second layer. So the surface atoms of the upper substrate and the fluid atoms of the first layer are moved together
a lattice distance to the right with respect to those
of the fourth layer adsorbed by the lower substrate. The slip processes with the multiple internal slips are also termed as the mixing slip.
For $T_{yy}=-1.0$, the slip process happens between the second and third layers in the film.
The fluid atoms of the second layer are moved a lattice distance to the right with respect to those of
the third layer.
For $T_{yy}=-1.5$, the slip process happens between the third and fourth layers in the film.
The fluid atoms of the third layer are moved a lattice distance to the right with respect to those of
the fourth layer.
So in the four-layer thin-film lubrication at the fixed $\kappa=1$ ($\epsilon_{sf}=1/3$),
the mixing slip is changed to the internal slip as the load increases.
In comparing with the effects of the film-substrate coupling strength on the system, it is found that
the effects of the load $-T_{yy}$ on the system  have similar tribological dynamical behaviors
(the thermal elastic constant and the width of the hysteresis loop),
but are opposite on the effective film thickness (or the mean separation $L_y$) in system.

\section{Conclusions and discussions}

Shearing transitions of multi-layer molecularly film lubrication systems
in variations of the film-substrate coupling strength and the load
have been studied by using a hybrid atomistic-coarse-grained method.
For the thin-film lubrication with multi-layer molecules at a fixed load,
three kinds of the interlayer slips (the internal, the interfacial and
the mixing slips) in the system are found as the film-substrate coupling strength decreases.
The internal and the interfacial slips have one relative slippage in the interlayers of the systems,
but the mixing slip has more relative slippages in them.
For the three-layer molecularly thin-film lubrication, the mixing slip includes the internal slip
and the upper and lower interfacial slips. For the four-layer molecularly thin-film lubrication,
the mixing slip includes the upper and lower interfacial slips or the multiple internal slips. 
These phenomena are in qualitative agreement with the experimental results\cite{a,b}.
Meanwhile, the tribological dynamical behaviors
(the thermal elastic constant and the width of the hysteresis loop)
are almost insensitive to the smaller film-substrate coupling strength.
However, they and the effective film thickness
are enlarged more and more as the larger film-substrate coupling strength increases.
When the load increases, the tribological dynamical behaviors are
similar to those in increasing film-substrate coupling strength, but the effective
film thickness is opposite.


\newpage
\textbf{Acknowledgments} This research is supported by the National Natural Science Foundation
of China through the Grants No. 11172310 and No. 11472284,
 and the CAS Strategic Priority Research Program XDB22040403.
The author thanks the IMECH research computing facility for
assisting in the computation.

\newpage
Table I. Parameters of the simulation. Reduced dimensionless units are based on the
Lennard-Jones parameters ($\sigma, \epsilon_{ss}$) of the substrate, as explained
in text.

\begin{tabular}{ll}
 \\ \hline
Number of fluid atoms for a threelayer film & $N_f= 30$
\\
Number of fluid atoms for a fourlayer film & $N_f= 40$
\\
Number of wall atoms & $N_w=2 \times 40 = 80$
\\
Total number of substrate atoms & $N_s=2 \times 240 = 480$
\\
Number of near-region atoms & $N'_s=2 \times 40 = 80$
\\
Number of far-region atoms & $N''_s=2 \times 200 = 400$
\\
Number of elements & $N_e=2 \times 4 = 8$
\\
Number of nodes & $N_n=2 \times 2 = 4$
\\
Lattice constant of substrate & $a=\sqrt{\frac{2}{\sqrt{3}}}=1.075$
\\
Width of contact & $L_x=10a$
\\
Temperature & $T=0.1$
\\
Substrate-substrate Lennard-Jones well depth & $\epsilon_{ss}=1$
\\
Film-film Lennard-Jones well depth & $\epsilon_{ff}=1/9$
\\
Cutoff radius & $r_c=2.5$
\\
Total number of Monte Carlo cycles for each $\alpha$ & $10^6$
\\ \hline
\end{tabular}

\newpage
FIGURE CAPTIONS

Fig.~1: Schematic of idealized two-dimensional contact with a thin fluid film for (a)
three-layer atoms/(b) four-layer atoms and partial coarse-graining of far regions of substrates.
  Filled triangles, opened squares and filled circles represent solid atoms in the walls, solid atoms in substrates
 and fluid atoms in the film, respectively.
 All atoms are depicted in their initial configuration at $\alpha=0$, where the atomistic positions
 in the film are random.

Fig.~2: (a) Shear stress $T_{yx,cf}$ versus register $\alpha$;
(b) mean-separation profile for the coarse-grained 2D model contact with the three-layer thin-film at a fixed load $T_{yy}=-1.0$.

Fig.~3: Atomistic structures of the fluid film and the near substrates
at the identical state at $\alpha=0$ for all $\kappa$ (a) and the different states at $\alpha=1.0$
for (b) $\kappa$ =1; (c) 1/3; (d) 1/6; (e) 1/12; (f) 1/16. $f$ and $s$ denote the fluid atoms in
the film and the solid atoms in the substrates, respectively.
The superscripts of $f$ and $s$ denote the layer coding in the film and the upper/lower substrates, respectively.
The subscripts of $f$ and $s$ denote the position coding in a layer at $\alpha=0$ in Fig.~3(a).

Fig.~4: (a) Shear stress $T_{yx,cf}$ versus register $\alpha$;
(b) mean-separation profile for the coarse-grained 2D model contact with the three-layer thin-film at a fixed
$\epsilon_{fs}=1/3$($\kappa$ =1).

Fig.~5: (a) Shear stress $T_{yx,cf}$ versus register $\alpha$;
(b) mean-separation profile for the coarse-grained 2D model contact with the four-layer thin-film at a fixed load $T_{yy}=-1.0$.

Fig.~6: Atomistic structures of the fluid films and the near substrates
at the identical state at $\alpha=-0.5$ for all $\kappa$ (a) and the different states at $\alpha=0.5$
for (b) $\kappa$ =1; (c) 1/6; (d) 1/9; (e) 1/12; (f) 1/20. Notation as in Fig.~3.

Fig.~7: (a) Shear stress $T_{yx,cf}$ versus register $\alpha$;
(b) mean-separation profile for the coarse-grained 2D model contact with the four-layer thin-film at a fixed
$\epsilon_{fs}=1/3$($\kappa$ =1).


\newpage
\begin{thebibliography}{50}
\bibitem{1}A. W. Adamson and A. P. Gast, Physical chemistry of surfaces, 6th ed., Wiley, New York, 1997.
\bibitem{2}M. Scherge and S. S. Gord, Biological mirco- and nanotribology, Springer, Berlin 2001.
\bibitem{3}M. Nosonovsky and B. Bhushan, Multiscale dissipative mechanisms and hierarchical surfaces, Springer, Berlin 2008.
\bibitem{3d}J. N. Israelachvili, P. M. McGuiggan and A. M. Homola, Dynamic properties of molecularly thin liquid films, Science {\bf 240}, pp. 189-91 (1988).
\bibitem{3e}J. V. Alsten and S. Granick, Molecular tribometry of ultrathin liquid films, Phys. Rev. Lett. {\bf 61}, pp. 2570-3 (1988).
\bibitem{ss}M. Schoen, C. L. Rhykerd, D. J. Diestler and J. H. Cushman, Shear forces in molecularly thin films, Science {\bf 245} pp.1223-5 (1989).
\bibitem{ds}D. J. Diestler, M. Schoen and J. H. Cushman, On the thermodynamic stability of confined thin films under shear, Science {\bf 262} pp.545-7 (1993).
\bibitem{a}J. N. Israelachvili, Intermolecular and surface force, 3rd ed., Academic Press, New York, 2011, pp.486-7.
\bibitem{b}S. Yamada, Nanotribology of symmetric and asymmetric liquid lubricants, Symmetry {\bf 2} pp. 320-45 (2010).
\bibitem{3a}J. Jiang, R. D. Arnell and J. Tong, The effect of substrate properties on tribological behaviur of composite DLC coatings, Tribol. Inter. {\bf 30}, pp. 613-25 (1997).
\bibitem{3b}J. M. Jungk, J. R. Micheal and S. V. Prasad, The role of substrate plasticity on the tribological behavior of diamond-like nanocomposite coatings, Acta Meta. {\bf 56},
pp. 1956-66 (2008).
\bibitem{3c}X. Wu, T. Ohana, T. Nakamura and A. Tanaka, Hardness effect of stainless steel substrates on tribological properties of water-lubricated DLC films against AISI 440C ball, Wear {\bf 268}, pp. 329-34 (2010).
\bibitem{4}S. Kohlhoff, P. Gumbsch and H.F. Fischmeister, Crack propagation in b.c.c crystals studies with a combined finite-element and atomistic model, Philos. Mag. A {\bf 64}, pp.851-78 (1991).
\bibitem{5}E. B. Tadmor, M. Ortiz and R. Phillips, Quasicontinuum analysis of defects in solids, Philos. Mag. A {\bf 73}, pp. 1529-63 (1996).
\bibitem{6}V. Shenoy, V. Shenoy and R. Phillips, Finite temperature quasicontinuum methods, Mater. Res. Soc. Sym. Proc., {\bf 538}, pp. 465-71 (1999).
\bibitem{6a}J. Q. Broughton, F. F. Abraham, N. Bernstein, and E. Kaxiras, Concurrent coupling of length scales: Methodology and application, Phys. Rev. B {\bf 60}, pp. 2391-403 (1999).
\bibitem{7}R. E. Rudd and J. Q. Broughton, Concurrent coupling of length scales in solid state systems, Phys. Status Solidi (b) {\bf 217}, pp. 251-91 (2000).
\bibitem{7a}M. Ortiz, A. M. Cuitino, J. Knap and M. Koslowski, Mixed atomistic continuum models of material behavior: The art of transcending atomistics and informing continua, MRS Bulletin, {\bf 26}, pp. 216-21 (2001).
\bibitem{8}R. E. Miller and E. B. Tadmor, The quasicontinuum method: Overview, applications and current directions, J. Computer-Aided Materials Design {\bf 9}, pp. 206-39 (2002).
\bibitem{9}R. M. Nieminen, From atomistic simulation towards multiscale modelling of materials, J. Phys.: Condens. Matter {\bf 14}, pp. 2859-76 (2002).
\bibitem{10}R. Phillips, M. Dittrich and K. Schulten, Quasicontinuum representations of atomic-scale mechanics: From proteins to dislocations, Annu. Rev. Mater. Res. {\bf 32}, pp. 219-33 (2002).
\bibitem{11}Z. B. Wu, D. J. Diestler, R. Feng and X. C. Zeng, Coarse-graining description of solid systems at nonzero temperature, J. Chem. Phys., {\bf 119}, pp. 8013-23 (2003).
\bibitem{11a}D. J. Diestler, Z.-B. Wu and X. C. Zeng, An extension of the quasicontinuum treatment of multiscale solid systems to nonzero temperature, J. Chem. Phys. {\bf 121}, pp. 9279-82 (2004).
\bibitem{12}M. Fago, R. L. Hayes, E. A. Carter and M. Ortiz, Density-functional-theory-based local quasicontinuum method: Prediction of dislocation nucleation, Phys. Rev. B, {\bf 70}, 100102 (2004).
\bibitem{13}R. E. Rudd and J. Q. Broughton, Coarse-grained molecular dynamics: Nonlinear finite elements and finite temperature, Phys. Rev. B, {\bf 72}, 144104 (2005).
\bibitem{14}S. P. Xiao and W. X. Yang, Temperature-related Cauchy-Born rule for multiscale modeling of crystalline solids, Comput. Mater. Sci. {\bf 37}, pp. 374-79 (2006).
\bibitem{15}V. Gavini, K. Bhattacharya and M. Ortiz, Quasi-continuum orbital-free density-functional theory: A route to multi-million atom non-periodic DFT calculation. J. Mech. Phys. Solids, {\bf 55}, pp. 697-718 (2007).
\bibitem{16}D. Negrut, M. Anitescu, A. El-Azab and P. Zapol, Quasicontinuum-like reduction of density functional theory calculations of nanostructures, J. Nanosci. Nanotech., {\bf 8}, pp. 3729-40 (2008).
\bibitem{17}N. Bernstein, J. R. Kermode and G. Csanyi, Hybrid atomistic simulation methods for materials systems, Rep. Prog. Phys. {\bf 72}, 026501 (2009).
\bibitem{18}J. Marian, G. Venturini, B. L. Hansen, J. Knap, M. Ortiz and G. H. Campbell, Finite-temperature extension of the quasicontinuum method using Langevin dynamics: entropy losses and analysis of errors, Model. Simul. Mater. Sci. Engin., {\bf 18}, 015003 (2010).
\bibitem{19}M. Iyer and V. Gavini, A field theoretical approach to the quasi-continuum method, J. Mech. Phys. Solids, {\bf 59}, pp. 1506-35 (2011).
\bibitem{20}X. H. Li, M. Luskin and C. Ortner, Positive definiteness of the blended force-based quasicontinuum method, Multiscale Model. Simul., {\bf 10}, pp. 1023-45 (2012).
\bibitem{20a}E. B. Tadmor, F. Legoll, W. K. Kim, L. M. Dupuy and R. E. Miller, Finite-temperature quasi-continuum, Appl. Mech. Rev. {\bf 65}, 010803 (2013).
\bibitem{21}W. K. Kim, M. Luskin, D. Perez, A. F. Voter and E. B. Tadmor, Hyper-QC: An accelerated finite-temperature quasicontinuum method using hyperdynamics, J. Mech. Phys. Solids, {\bf 63}, pp. 94-112 (2014).
\bibitem{22}J. S. Amelang, G. N. Venturini and D. M. Kochmann, Summation rules for a fully-nonlocal energy-based quasicontinuum method, J. Mech. Phys. Solids, {\bf 82}, pp. 378-413 (2015).
\bibitem{16a} W. G. Noid, J.-W. Chu, G. S. Ayton, V. Krishna, S. Izvekov, G. A. Voth, A. Das and H. C. Anderson,
The multiscale coarse-graining method. I. A rigorous bridge between atomistic and coarse-grained models,
J. Chme. Phys. {\bf 128}, 244114 (2008).
\bibitem{16b} W. G. Noid, P. Liu, Y. Wang, J.-W. Chu, G. S. Ayton, S. Izvekov, H. C. Anderson and G. A. Voth,
The multiscale coarse-grainning method. II. Numerical implementation for coarse-grained molecular models,
J. Chem. Phys. {\bf 128}, 244115 (2008).
\bibitem{16c} P. Sherwood, B. R. Brooks and M. S. Sansom, Multiscale methods for macromolecular simulations,
Curr. Opin. Struct. Biol. {\bf 18}, 630 (2008).
\bibitem{19a} H. L. Woodcock, B. T. Miller, M. Hodoscek, A. Okur, J. D. Larkin, J. W. Ponder and B. R. Brooks,
MSCALE: a general unility for multiscale modeling, J. Chem. Theory Comput. {\bf 7}, 1208 (2011).
\bibitem{18a} A. Singharoy, A. M. Yesnik and P. Ortoleva, Multiscale analytic continuation approach to
nanosystem simulation: Application to virus electrostatics, J. Chem. Phys. {\bf 132}, 174112 (2010).
\bibitem{20a} A. Singharoy, H. Joshi and P. J. Ortoleva, Multiscale macromolecular simulation: role of evolving
ensembles, J. Chem. Inf. Model. {\bf 52}, pp.2638-2649 (2012).
\bibitem{24}Z.-B. Wu, D. J. Diestler and X. C. Zeng, Multiscale simulation of thin-film lubrication, Mol. Simul. {\bf 31}, pp. 811-5 (2005).
\bibitem{25}Z.-B. Wu and X. C. Zeng, Multiscale simulation of thin-film lubrication: Free-energy corrected coarse graning, Phys. Rev. E {\bf 90}, 033303 (2014).
\bibitem{c}O. M. Braun, M. Paliy and S. Consta, Ordering of a thin lubricant film due to sliding, Phys. Rev. Lett. {\bf 92}, 256103 (2004).
\bibitem{lesar}R. LeSar, R. Najafabadi, and D. J. Srolovitz, ``Finite-temperature defect properties from free-energy minimization'', Phys.  Rev. Lett. {\bf 63}, pp. 624-7 (1989).
\bibitem{26}D. J. Diestler, G. T. Gao and X. C. Zeng, Role of hyperesis in the molecular picutre of friction, Phys. Chem. Chem. Phys. {\bf 3}, pp. 1175-8 (2001).
\bibitem{27}D. J. Diestler, Constranied statistical thermodynamic treatment of friction, J. Chem. Phys. {\bf 117}, pp. 3411-24 (2002).
\end{thebibliography}
\end{document}